\documentclass[12pt]{iopart}
\usepackage{latexsym}
\usepackage{bm}
\usepackage{slashed}
\usepackage{graphicx}
\newcommand*{\zed}[1]{\bm{\mathrm{Z}}_{{#1}}}

\newcommand*{\vect}[1]{\bm{{#1}}}

\newcommand*{\vev}[1]{\langle{{#1}}\rangle}

\newcommand*{\complex}[1]{\bm{\mathrm{C}}^{{#1}}}
\newcommand*{\order}[1]{\mathcal{O}({#1})}
\newcommand*{\imag}{\mathrm{i}}
\newcommand*{\scale}{\ell}

\newcommand*{\planckmass}{M_{\stext{P}}}
\newcommand*{\susymass}{M_{\stext{\textsc{susy}}}}
\newcommand*{\mcutoff}{M_{\stext{cut-off}}}
\newcommand*{\vloop}{V^{\stext{L}}}
\newcommand*{\aloop}{\mathcal{A}_{\stext{L}}}
\newcommand*{\atri}{\mathcal{A}_{\stext{T}}}
\newcommand*{\pout}{p_{\stext{out}}}
\newcommand*{\qclass}{Q_{\stext{cl}}}

\newcommand*{\yh}{y_{\stext{h}}}
\newcommand*{\iso}{\bm{\mathrm{R}}^3}
\newcommand*{\gutmass}{M_{\stext{\textsc{gut}}}}
\newcommand*{\euclideank}{k_{\stext{E}}}
\newcommand*{\diag}{\mathrm{diag}\,}

\newcommand*{\pd}[2]{\frac{\partial{{#1}}}{\partial{{#2}}}}
\newcommand*{\ptwo}[2]{\frac{\partial^2{{#1}}}{\partial{{#2}}^2}}

\newcommand*{\measure}[1]{\mathrm{d}{#1}}

\newcommand*{\fmeasure}[1]{[\measure{{#1}}]}
\newcommand*{\dthree}[1]{\mathrm{d}^3\vect{{#1}}}
\newcommand*{\dfour}[1]{\mathrm{d}^4{#1}}
\newcommand*{\dfive}[1]{\mathrm{d}^5{#1}}

\newcommand*{\laplacian}{\triangle}

\newcommand*{\bop}[1]{\mathcal{B}_{{#1}}}
\renewcommand{\e}[1]{\mathrm{e}^{{#1}}}

\newcommand{\eqref}[1]{Eq.~(\protect\ref{#1})}
\renewcommand{\epsilon}{\varepsilon}
\newcommand{\stext}[1]{\mbox{\scriptsize{{#1}}}}
\newcommand{\ntext}[1]{\mbox{{#1}}}
\providecommand*{\etal}{\emph{et al.}}
\begin{document}
\title{Radiative constraints on brane quintessence}
\date{\today}
\author{David Seery\dag, Bruce A. Bassett\ddag\footnote[3]{%
Permanent address: Institute of Cosmology and Gravitation,
University of Portsmouth, Mercantile House, Portsmouth, PO1 2EG,
United Kingdom}}
\address{\dag\ Institute for Astronomy, The University of Edinburgh,
Blackford Hill, Edinburgh, EH9 3HJ, United Kingdom}
\address{\ddag\ Department of Physics, Kyoto University, Kyoto,
Japan}
\ead{\mailto{djs@roe.ac.uk}, \mailto{bruce@tap.scphys.kyoto-u.ac.jp}}
\submitto{JCAP}
%
%
\begin{abstract}
We investigate the constraints on quintessence arising from both renormalisable
and non-renormalisable couplings where the 5d Planck mass is around the TeV scale.
The quintessence field vacuum expectation value is typically of order the 4d Planck
mass, while non-renormalisable operators are expected to be suppressed by the 5d
Planck mass. Non-renormalisable operators are therefore important in computing the
4d effective quintessence potential. We then study the quantum corrections to the
quintessence potential due to fermion and graviton loops. The tower of Kaluza--Klein
modes competes with the TeV-scale cut-off, altering the graviton contribution to the
vacuum polarization of quintessence. Nevertheless we show that, as in four dimensions,
the classical potential is stable to such radiative corrections. \\[2mm]
\begin{flushleft} Keywords: dark energy theory, cosmology with extra dimensions \\
astro-ph/0310208 \end{flushleft}
\end{abstract}
%
\maketitle

%
%
\section{Introduction}
The case for the existence of a $\Lambda$-like component
dominating the currently observable universe is now
compelling \cite{wmap}. In the simplest models such a component has two quite independent
properties. On the one hand, it does not cluster on scales much smaller than the 
Hubble scale, $H^{-1}$, and on the other it influences the background evolution of the 
cosmos, causing acceleration at very recent redshifts and giving rise to the coincidence 
problem -- why do we appear to be living at a special time in the Universe's history? 
While it is possible to construct models which exhibit only one of these properties, eg.
\cite{condens, axionphoton}, both now have observational support. In particular, the recent
detection of cross-correlations between the Wilkinson Microwave Anisotropy Probe (WMAP) cosmic
microwave background (CMB) anisotropies and various tracers of 
large scale structure \cite{boughn,wmap,scranton,fosalba}, which are consistent with the decay of perturbations
on large scales due to an accelerating background, make construction of convincing non-accelerating
models difficult. 

However, despite its success at the purely phenomenological level, the standard $\Lambda$CDM model  
has almost no deep understanding to back it up. We are in the age of precision 
book-keeping in cosmology, but despite many attempts we do not yet have even a well-founded
theoretical order-of-magnitude estimate of the size of the
cosmological constant: most na\"{\i}ve field theory calculations
disagree by $\order{10^{120}}$ with observations, yielding perhaps the worst estimate in 
the history of physics. One can improve the situation somewhat by invoking supersymmetry, but it
proves generically quite hard to construct supergravity vacua
with positive cosmological constant \cite{townsend}.  
The string theory case is even harder
\cite{stringy-desitter}.  In the absence of any theoretical control
over $\Lambda$ itself, there is a strong temptation to explain the observations
by invoking some other mechanism.  Some proposals utilise the large number of possible
string theory vacua, either by appealing to the anthropic principle
\cite{susskind} or other quantum effects \cite{kane-perry},
but a more moderate approach is simply to include, among the matter inventory of the universe,
some tensile matter with appropriate equation of state whose behaviour is under good control.
This allows us to set $\Lambda = 0$ by supposing that one or more of the string theory proposals
for cancelling $\Lambda$
applies, and then to exclude the complexities of $\Lambda$ itself and deal instead with
the relatively well-understood properties of matter.
We will argue that, at least in TeV-scale models of quintessence, control is 
not manifest even in this case. 

Quintessence consists of a scalar field $Q$, which drives a late-time accelerated cosmological 
expansion via its vacuum expectation value in a rather similar way to scalar-field driven inflation.
If $Q$ is still rolling today then it must be very light in order to satisfy the 
standard slow-roll conditions and hence its 
Compton wavelength, $\lambda_c \simeq V_{QQ}^{-1/2}$, is very large (we denote first, second, \ldots,
$Q$-derivatives of $V$ as $V_Q$, $V_{QQ}$, \ldots, etc.)  As a result 
it only clusters on very large scales, typically greater than $100$ Mpc. 

This is not necessary.  In models involving the Albrecht--Skordis potentials, where 
the dark energy reaches a minimum of the potential at non-zero energy, the mass and expectation 
value of the quintessence is arbitrary, and so the dark energy may cluster on all scales after reaching the 
minimum. Such models are attractive for another important reason, for if the quintessence
is very light (it typically has a mass $m_{Q} \sim 10^{-33} \; \ntext{eV}$), then we must find a 
way to protect this mass from radiative corrections which will otherwise spoil the 
flatness of the potential \cite{rad-a} (see also, eg., Refs. \cite{rad-b,
pilo-rayner}).
This has been studied at the one-loop level
\cite{doran-jackel}.  The result depends on which particle
species one includes in the loops.  One typically finds
that couplings to bosons are benign \cite{doran-jackel},
but couplings to fermions are severely constrained.  The bounds found by the
authors of Ref. \cite{doran-jackel} are extremely stringent
and give rise to concern that gravitational couplings alone
might be strong enough to violate them.  Estimates presented in Ref. \cite{doran-jackel}
show that quintessence is safe, but this safety is model-dependent and must
be assessed carefully.

In addition, $Q$ must be extremely weakly coupled to standard model 
fields, otherwise it is difficult to see how it could have evaded detection via
particle physics or cosmological interactions.  Despite their overall weakness, such couplings can alter standard 
cosmology in an interesting way \cite{amendola-a,amendola-b,maccio}, but 
obtaining them requires significant fine-tuning of the renormalisable couplings. 

A more worrying problem is provided by constraints from 4d non-renormalisable couplings between the 
standard model and the quintessence field. Such couplings are generically expected from
supergravity and string 
theory, and are problematic in `tracking' quintessence models which generally have Planckian vacuum expection
values (vevs). Couplings such as $\beta Q F^2/M$, where $F^2$ is the usual Maxwell Lagrangian and 
$M$ is the mass-scale at which we expect supergravity to fail as an effective theory cause variation of 
the fine-structure constant, and because of the large $Q$-vev require fine-tuning of the dimensionless coupling
$\beta$ of order $\beta < 10^{-5}$. Since we have no reason to expect $\beta$ to differ significantly from order unity,
this unexplained fine-tuning is unsettling. Carroll \cite{carroll} has argued that such dimension-five
operators may be excluded by the existence of a discrete $\zed{2}$ symmetry in the fundamental description, which
acts on the extra dimension as $\phi \rightarrow -\phi$, but 
even in this case such fine-tuning persists with higher-order operators of the form $Q^n F^2/M^n$ \cite{PBB}.
One of our aims is to consider the effect of such non-renormalisable couplings in models with a low-scale
of quantum gravity. 

There are many constraints one can consider. 
Despite arising from a variety of different physics, these bounds and the constraints
on fermion couplings arising from stability of the
classical potential share a common feature: they are sensitive to some power of
the ratio $M/\mcutoff$, where $M$ is some energy scale characteristic of the process in question,
and $\mcutoff$ is an energy scale which controls the details
of heavy physics, which we consider to have been integrated out in
our effective description.

In the conventional kind of four-dimensional
cosmology one would usually take $\mcutoff$ to be
of order the Planck mass $\planckmass = G^{-2} \approx 10^{19} \; \ntext{GeV}$,
although there are other natural candidates: the \textsc{gut} scale
$\gutmass$ around $10^{15} \; \ntext{GeV}$; the
string scale, possibly a few orders of magnitude less than the
Planck mass; or, more speculatively, the \textsc{susy} scale,
which may be as low as $\susymass \sim 100 \; \ntext{GeV}$.
In recent years, inspired by ideas from the strongly coupled limit
of the heterotic string \cite{horava-witten-a, lukas-a, lukas-b,
randall-sundrum-A, randall-sundrum-B}
an alternative scenario for cosmology has become popular,
in which the various gauge and matter fields which comprise our
universe are affixed to a hypersurface in
a larger, five-dimensional bulk spacetime.  This spacetime is generically
a patch of Schwarzschild--anti de Sitter (SAdS) space.  In these
models, the fundamental scale $\planckmass$ of quantum gravity might be
much lower, perhaps only of order a TeV ($10^{-16} \, \planckmass
= 10^{12}$ eV) or so, in which case
one would expect to obtain very different constraints on
quintessence.  (However one should note that in these models, one cannot really be
dealing with the heterotic string, since in such theories the string scale---which amounts
to the quantum gravity scale---is fixed
to roughly coincide with the four-dimensional Planck scale and there is not much freedom
to move it.  On the other hand, theories such as Type I string theory can acceptably
accommodate a low string scale.)
Because the troubling bounds and fine-tunings outlined above depend sensitively on the
details of the cutoff scale, one should carefully recalculate the constraints they impose
in models with quantum gravity at low energies, but this is not sufficient.
There are other effects which one should also take into account, arising from new
physics associated with the branes.  Most notably, for example,
these models contain a tower of Kaluza--Klein modes in addition to a massless four-dimensional
graviton.
These Kaluza--Klein modes can be considered to arise from gravity in the bulk.
These modes are typically massive, with masses $m > 3H/2$
where $H$ is the brane Hubble parameter \cite{lmw,gorbunov,frolov-kofman}.
The presence of these modes
introduces processes, absent in four dimensions, where quintessence can interact with gravity off
the brane, or with the Kaluza--Klein hierarchy.

In this paper, we apply all these ideas to constrain quintessence
couplings and energy scales in TeV scale Planck mass models. Higher dimensional 
and brane-world models offer interesting new insights into quintessence cosmology
\cite{mizuno-a,mizuno-b,burgess,hill,albrecht-burgess}, but before adopting these
models wholesale it is important to consider potential constraints and compare 
them with the corresponding constraints on 4d quintessence
\cite{doran-jackel,horvat,carroll}.

The plan of this paper is as follows.  In the next section we discuss the 
issue of non-renormalisable couplings.  Then we briefly review
the bounds on quintessence couplings in four dimensions, paying
particular attention to bounds on couplings to fermions.
In Section~\ref{sec:grav-couple} we calculate probability amplitudes for
some representative gravitational processes, where bulk gravitons mediate quintessence
couplings to fermions.  We also estimate the lowest-order
contribution of virtual graviton exchange to the vacuum polarization
$\Pi^\ast(p)$ of the quintessence field.  In Section~\ref{sec:conclusions}
we state our conclusions.
In an appendix, we give a brief derivation of the
gravitational propagator in the brane world, using the Fadeev--Popov technique.
This has appeared in the literature before \cite{giddings-katz} but we present
this alternative derivation for simplicity and to make our account self-contained.

Throughout we work in units where $\hbar = 1$, but the gravitational
coupling in $D$ dimensions is $\kappa_D^2 = 8\pi / \planckmass^{D-2}$,
where $\planckmass$ is the $D$-dimensional Planck mass.  We use
eV as units of dimensionful quantities everywhere.

%
%
\section{Non-renormalisable couplings between quintessence and the standard model}
One of the original motivations for TeV-scale quantum gravity was its ability to obviate the 
need for low-energy supersymmetry by removing the huge hierarchy between the weak and Planck scales. 
Such a low scale for quantum gravity can be achieved rather elegantly in models of large
extra dimensions \cite{antoniadis,dvali} or in brane-world models with a non-compact extra dimension. 

Viewed from this perspective, the four-dimensional theory on the brane is simply an effective theory
arising from the dimensional reduction of a more fundamental, higher-dimensional theory by integrating 
out physics at scales above which the extra dimensions become visible. As such, we must expect our 
effective four-dimensional theory to be burdened with a potentially infinite number of non-renormalisable 
interactions, suppressed by powers of the mass scale at which the effective theory breaks down or at which new physics enters the problem.
In this case it is natural to assume the cutoff scale $\mcutoff$
to be the Planck scale of the higher-dimensional
theory, $\mcutoff \sim \planckmass \sim \ntext{TeV}$ provided that this
is lower than the effective 4d Planck scale.  Above the scale $\mcutoff$ the 
theory is no longer well-approximated by a four-dimensional theory. 

In the discussion that follows we will leave the cutoff mass scale $\mcutoff$ arbitrary.
To minimise clutter, we write this scale as $M$. Our conclusions are strongest in models where $M \sim \ntext{TeV} $ and weaken as $M$ increases towards the four dimensional value $\planckmass \sim 10^{16} \; \ntext{TeV}$. 

Let us begin by considering 
non-renormalisable Lagrangian operators of the form (see eg. \cite{carroll})
\begin{equation}
  \beta_n \frac{Q^n}{M^n} {\cal L}_4
\end{equation}
where ${\cal L}_4$ is any dimension-four standard-model operator such as $F^2$ or 
$G^2$.  We are seeking constraints on
the dimensionless coupling $\beta$, and in particular its dependence on the cut-off scale, as discussed
above.
The case ${\cal L}_4 = F^2$ leads to cosmic variation of the fine-structure constant $\alpha$ \cite{carroll, PBB} which, assuming slow variation of $Q$, gives
\begin{equation}
  \Delta \alpha \simeq -n \beta_n Q^{n-1}|_{\overline{Q}} \frac{\Delta Q}{M^n}
\label{alpha}
\end{equation}
where $\overline{Q} = Q(0)$; the symbol $\Delta Q$ abbreviates the field interval
$\Delta Q(z) = Q(z) - Q(0)$; and $z$ denotes redshift.
The Webb \etal results suggest a variation of
$\alpha$ at the level $\frac{\Delta \alpha(z)}{\alpha} = (-0.543\pm 0.116) \times 10^{-5}$, ie.
evidence for variation of $\alpha$ at the $4.7\sigma$ level \cite{webb}. For $n \geq 1$ the ratio
$Q/M$ clearly plays a key role. We have argued that in TeV-scale higher dimensional models $M$
may be close to the TeV scale. The estimate of the vev $\vev{Q}$ is model dependent; however,
we can obtain constraints on standard tracker quintessence models where the dark energy equation
of state tracks that of the dominant energy component of the Universe until a low redshift
$z \sim 1$. In such models the quintessence field has been rolling since very early times and it
is natural that it has a large vev today, whether the fundamental theory be four- or higher dimensional. 

We can get a rough lower-bound on the vev of $Q$ today by the following argument. To ensure that the 
Universe accelerates, the field $Q$ must satisfy the standard slow-roll conditions familiar 
from early universe inflation. That is, we demand $\epsilon, \eta \ll 1$ where $\epsilon \equiv
\planckmass^2 (V_Q/V)^2$ and $\eta \equiv \planckmass^2 (V_{QQ}/V)$. At this point it is appropriate
to comment on the mass scale appearing in $\epsilon$ and $\eta$ and whether we really should be
using the four-dimensional Planck mass $\planckmass$, or the fundamental scale $M$ \cite{braneinflation}.
To answer this one can examine the effective four-dimensional equation of motion for $Q$. Slow-roll
requires that the potential term $V_{Q}$ be sub-dominant with respect to the friction term coming
from the Hubble expansion.  At least at low energies, the effective Friedmann equation giving the
Hubble expansion typically contains $\planckmass$, not $M$, and hence the effective four-dimensional
scale $\planckmass$ is the appropriate scale.

To quantify the constraints coming from requiring $\epsilon, \eta \ll 1$, consider a standard tracking
potential $V(Q)\propto Q^{-\gamma}$. In this case the field evolves as $Q \propto t^{2/(2+\gamma)}$
and the slow-roll parameter $\eta$ is
\begin{equation}
\epsilon = \gamma^2 \frac{\planckmass^2}{Q^2} \Longrightarrow Q \gg \gamma \planckmass
\end{equation}
with a similar constraint arising from $\eta \ll 1$.  Clearly one can make the vev as small as one
likes by  fine-tuning the exponent $\gamma$ to be sufficiently small but then one loses the
attractiveness of the model, since  one is simply converging to a cosmological constant. Even then
the fine-tuning required is significant. To have $\eta \leq 1$ with $Q \leq \ntext{TeV}$ requires
$\gamma \leq 10^{-16}$.  This is quite undesirable.

With this restriction on the form of the potential in mind, let us return to the dimensionless couplings
$\beta$.  If we use the standard result that tracking models typically require vevs of order the 4d Planck
scale, $Q \sim \planckmass$, then this implies that the ratio $Q/M$ can be as large as
$\planckmass/\ntext{TeV} \sim 10^{16}$.  Matching the Webb \etal data for $n = 1$ is possible with either a
fine-tuning of $\beta \sim 10^{-5}$ (the result of \cite{carroll}) or is compatible with
$\beta \sim 1$ by requiring $\Delta Q \sim 10^{-5} M = 10 \; \ntext{MeV}$.  Such a slow variation of the
field is not consistent with the assumption of a rolling quintessence.  However it is consistent with the
Albrecht--Skordis \cite{AS} model where $Q$ becomes trapped at the minimum of a potential well.
Interestingly, a more detailed analysis shows that the union of constraints on $\alpha$ at various
redshifts favour very little variation of $Q$ at low redshifts $z < 2$ even when $M = \planckmass$ \cite{PBB}.
 
For $n > 1$ the situation is much worse, since the large ratio $(Q/M)$ appears. The required
fine-tuning on the dimensionless couplings $\beta_n$, or variation $\Delta Q$, become enormous.
For $n = 2$ we have $\beta_2 \Delta Q \sim 10^{-9} \; \ntext{eV}$, or if we conservatively set
$\Delta Q \sim M$ we find $\beta_2 \sim 10^{-21}$.  Such a coupling is quite inexplicably small.
As $n$ increases  the fine-tuning on the dimensionless couplings $\beta_n$ rapidly increases.
Of course this simply underlines a more fundamental point: since $(Q/M) \gg 1$ we have absolutely
no control over the effective potential of the quintessence field. The potential should be computed
from the higher-dimensional theory. This is a standard argument against chaotic inflation which
typically requires super-Planckian initial conditions to obtain sufficient inflation and the
correct amplitude of density perturbations. 

Of course, one may simply argue that the non-perturbative potential for $Q$ is completely
well-behaved and only gives rise to small couplings to the standard model fields. This is
reminiscent of the runaway dilaton model of quintessence \cite{runaway} in which a massless
dilaton runs to infinity where it decouples from all matter, as in the proposal of
Damour \& Polyakov \cite{DP}. If the quintessence field is a radion, representing the distance between
two branes, driven apart, for example, by a repulsive Casimir force, then we
expect the $Q/M \rightarrow \infty$ limit to be trivial: the potential
should vanish.  This occurs because the $Q$ field dynamics effectively vanish in this limit:
it is equivalent to the degenerate case where the branes approach each other.
Couplings to gauge fields localised on the brane world
could also reasonably be expected to vanish in that limit.  However,
to appreciate this one requires the full five-dimensional picture, and in
other scenarios the corrections may not be so harmless.

%
%
\section{Quintessence couplings in four dimensions}
In the above analysis, we obtained restrictions on a given set of coupling
constants and shape parameters for some popular, rather generic potentials.
In doing so we assumed that the form of the potential could simply be given
as an Ansatz, so strictly speaking we were dealing with renormalized quantities,
and the potential was the quantum effective potential.  A more subtle question
is to ask how a given tree-level potential may be modified when quantum effects
are taken into account.  This involves the study
of loop corrections to the quintessence potential 
and the couplings of $Q$ to
other fields.

For scalar quintessence and fermions, this was first done by Doran \& J\"{a}ckel 
\cite{doran-jackel} and by Horvat \cite{horvat}, who considered
couplings to neutrinos.
We review their arguments as applied in four dimensions, and explain
how this generalizes to the brane world.  Many of the bounds
described in Ref. \cite{horvat} do not depend on the scale of gravity
and are not strongly modified in the brane world, so we focus on the
gravitational couplings described in Ref. \cite{doran-jackel}.
In particular, we are
interested in the coupling of quintessence to fermions, for which the
strongest bounds apply.

\begin{figure}
\begin{center}
\includegraphics[clip=true]{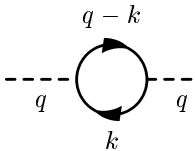}
\end{center}
\caption{Lowest order fermion loop contributing to effective quintessence
potential.  The dashed lines represent quintessence particles, and
solid lines are fermions.\label{fig:fermion-loop}}
\end{figure}

In the quintessence sector we work with the Euclidean action
\begin{equation}
  \label{eq:fourd-matter}
  S = \int \dfour{x} \; \sqrt{-g} \left( \frac{1}{2} \partial_a Q \partial^a
  Q + V(Q) + \bar{\psi}(\slashed{\partial} + m(Q))\psi \right)
\end{equation}
where $Q$ is the quintessence field, $V(Q)$ is its
classical potential, $\psi$ is a Dirac fermion,
and $m(Q)$ is a possibly $Q$-dependent fermion mass.
The leading
correction to the quintessence potential in the fermionic sector
comes from the diagram of Fig. \ref{fig:fermion-loop}.

Let $m$ be given by a large field independent mass $m_0$ plus some
correction $c$ generated by couplings to other fields.  Then the
condition that the classical potential dominates becomes
\cite{doran-jackel}
\begin{equation}
  \label{eq:fermion-bound}
  \frac{\vloop}{V} = \frac{1}{4\pi^2} \frac{\Lambda^2 m_0 c}{V} \ll 1 ,
\end{equation}
where we have discarded the $c$-independent piece $m_0^2$ which does not
affect dynamics, and we are assuming that $c \ll m_0$.
One can estimate $V$ \cite{doran-jackel}
by supposing that $Q$ currently dominates the
energy density of the universe, so that $V$ must be comparable to
$\rho_{\stext{crit}} = 8.1 \times 10^{-11} \; h^2 \; \ntext{eV}^4$.
Setting $\Lambda$ at around the \textsc{gut} scale
$\Lambda = 10^{-3} \; \planckmass$, and taking the
field-independent fermion mass to be around the supersymmetry breaking
scale, perhaps of order a TeV, or $m_0 = 10^{-16}
\; \planckmass$, that gives
\begin{equation}
  \label{eq:c-bound}
  c \ll 10^{-71} \; \ntext{eV} .
\end{equation}

This calculation only depends on the details of quantum field theory
in the four-dimensional world, so it is valid on the world volume
of a brane universe provided that we take the effective cut-off
$\Lambda$ to be sized appropriately.  Since the bound on $c$ scales
with $\Lambda^{-2}$, this means that a reduced cut-off will weaken
any constraint.  For example,
in a model where $\Lambda$
should be $\order{\ntext{TeV}} \sim 10^{12} \; \ntext{eV}$, one finds
$c \ll 10^{-44}$ eV.  This weakening is a mixed blessing.  It is
harder to rule out any given quintessence model, but it may make the
construction of a viable phenomenological model easier.

%
%
\section{Gravitationally mediated couplings in the brane world}
\label{sec:grav-couple}
\begin{figure}
\begin{center}
\includegraphics[clip=true]{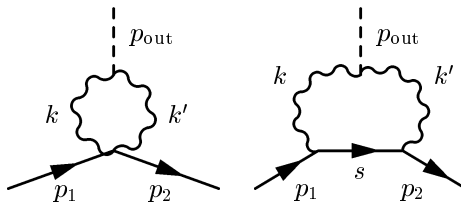}
\end{center}
\caption{Low order gravitationally mediated diagrams contributing to the
fermion--quintessence coupling.  Fermions are solid lines; gravitons are
wavy lines.  Quintessence is shown as dashes, and since the graviton
couples to the entire quintessence potential, this in principle
involves arbitrary powers of $Q$.\label{fig:grav-couple}}
\end{figure}

This bound \eqref{eq:c-bound}, in brane or four-dimensional form, is rather stringent and
could conceivably be violated by gravitational couplings.
The low order diagrams showing the gravitational coupling of
quintessence to fermions are shown in Fig. \ref{fig:grav-couple}.
In four dimensions,
both of these diagrams involve the classical quintessence
potential, so in fact the bound \eqref{eq:c-bound} does not apply.
This happens because one can absorb the corrections into a renormalization of
$V(Q)$ \cite{doran-jackel}.  We will compute the brane world
diagrams equivalent to Fig. \ref{fig:grav-couple}.

The principal result we shall require is the propagator for gravitons
in the bulk.  As a first approximation, we work in the original Randall--Sundrum
scenario where the branes are exactly flat.  The metric is
\begin{equation}
  \measure{s}^2 = \frac{1}{\scale^2 z^2} \left( -\measure{t}^2 + \measure{\vect{x}}^2 +
  \measure{z}^2 \right) .
\end{equation}
The propagator has been derived by Giddings, Katz \& Randall
\cite{giddings-katz}, and is given by
\begin{equation}
  \left .\Delta^{rsmn}\right|_{\stext{on-brane}} =
  \frac{2 \kappa^2}{\scale R} \rho^{rsmn} \int
  \frac{\dfour{k}}{(2\pi)^4} \e{-i k \cdot (x-y)} \frac{1}{k}
  \frac{K_2(kR)}{K_1(kR)}
\end{equation}
where $K_\nu$ is the Macdonald or Basset function
\begin{equation}
  K_\nu(z) = \frac{1}{2} \pi i^{\nu + 1} H_\nu^{(1)}(iz) ,
\end{equation}
the brane is fixed at $z = R$ (for RS branes, $R = \scale^{-1}$),
and $\rho^{rsmn}$ satisfies
\begin{equation}
  \rho^{rsmn} = \delta^{r(m} \delta^{n)s} - \frac{1}{d-2} \delta^{rs}
  \delta^{mn} ,
\end{equation}
where $d = \delta^i_i$ is the trace of the $\iso$ Kronecker delta.
This is 3 in the present case but changes if one sends $\iso$ to
$\mathbf{R}^n$.
We give an alternative derivation of this result in an appendix.

It is convenient to use the Feynman rules in configuration space,
rather than momentum space.  This is because the brane matter
theory has support only on the brane and its couplings
naturally include
a term $\delta(z - R)$ which is most easily accommodated in the
configuration space formulation.  The presence of the brane breaks
bulk translational isometries in the transverse direction, and there
is no conserved Noether charge to play the role of a conserved
momentum in the $z$ direction.  For this reason, loop diagrams will still
involve an integration over four-momenta on slices $z = \ntext{constant}$,
and not full five-momenta in the bulk.

\begin{figure}
\begin{center}
\includegraphics[clip=true]{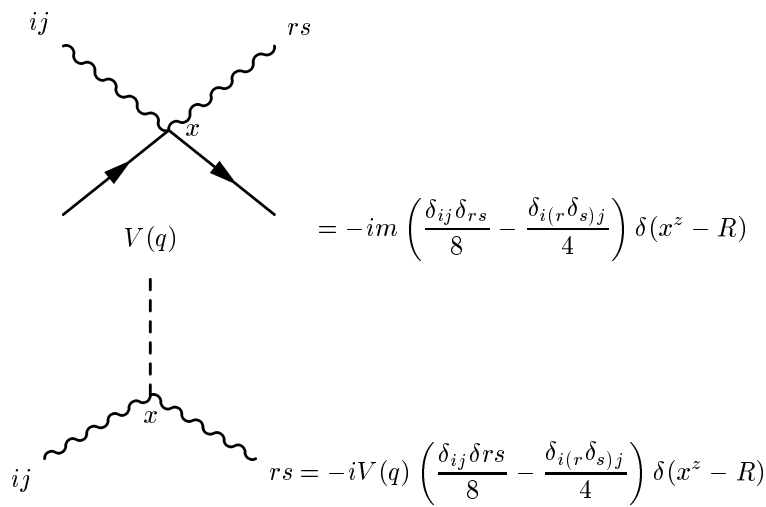}
\end{center}
\caption{Some vertices in the quantum field theory corresponding to
\eqref{eq:fived-matter},
where Dirac indices are suppressed.
Fermions are solid lines.  Gravitons are indicated by wavy lines,
and quintessence by dashes.  The vertices are written in configuration
space and taken to occur at a spacetime point $x$.  Indices
$ij$, $rs$ on graviton lines refer to the $\iso$ index structure.
Fermions entering the diagram at a point $x$
with momentum $p$ carry a
coefficient function $(2\pi)^{-3/2} u(p) \e{i p \cdot x}$,
where $\cdot$ denotes the flat, Euclidean inner product on the
brane.  Fermions leaving the
diagram from a point $x$ with momentum $p$ carry $(2\pi)^{3/2} \bar{u}(p)
\e{-i p \cdot x}$.  Here, the constant functions $u$, $\bar{u}$
are Dirac spinors with indices suppressed: our conventions for
spinors and $\gamma$-matrices match Weinberg \cite{weinberg}.
Quintessence particles entering or leaving
the diagram with momentum $p$ carry $(2\pi)^{-3/2} \e{\pm i p \cdot x}$.
To find amplitudes, one integrates over the coordinates $x_1$,
$x_2$, \ldots of all interaction points.  These integrals should
include the appropriate volume measure $\sqrt{-\det g}$, which is
unity on the brane.  In
addition, scalar products are taken in the metric
$h_{ab}$ with is the pull-back of the spacetime metric $g_{ab}$ to
the slice $z = \ntext{constant}$ over which they are evaluated;
this reduces to the flat, Euclidan scalar product on the brane.
\label{fig:vertices}}
\end{figure}

The on-brane
gravitational propagator is as given in \eqref{eq:on-brane-prop},
and the fermion and scalar propagators are as usual.
We introduce a matter theory on the brane described by the
analogue of the four-dimensional quintessence--fermion system
\begin{equation}
  \label{eq:fived-matter}
  S_{\stext{brane}} = \int_{z = R} \dfour{x} \sqrt{-\det h} \;
  \left[ \frac{1}{2} \partial_a Q \partial^a Q + V(Q) + \bar{\psi}
  ( -i \slashed{\partial} + m ) \psi \right]
\end{equation}
where, $h_{ab}$ is the pull-back of the five dimensional metric $g_{ab}$
to the brane, $Q$ is the quintessence field and $\psi$ is a
four-dimensional (not five-dimensional) Dirac fermion.
We are ignoring any gravitational coupling to $\psi$ via the
spin connexion.
In the Randall--Sundum case, $h_{ab}$ is just
four-dimensional Minkowski space plus the tensor perturbation
$e_{ij}$ evaluated at $z = R$, so
\begin{equation}
  \sqrt{-\det h} = 1 + \frac{\tr e}{2} - \frac{\tr e^2}{4}
  + \frac{(\tr e)^2}{8} + \cdots
\end{equation}
The vertices for this theory are shown in Fig. \ref{fig:vertices}.
(See also, eg., Ref. \cite{han-lykken}.)

We can now proceed to evaluate the diagrams
in Fig. \ref{fig:grav-couple} with the brane world graviton propagator.

\subsection{Loop diagram}
Consider the first diagram in Fig. \ref{fig:grav-couple}.
The amplitude for this process is
\begin{equation}
  \fl
  \aloop = \frac{5i}{2} \frac{m \kappa^2}{\scale^2 R^2} V(\qclass)
  \frac{\bar{u}(p_1) u(p_2)}{(2\pi)^{9/2}} \delta(\sum p)
  \int
  \dfour{\euclideank} \, \left . \frac{1}{k} \frac{1}{k'} \frac{K_2(kR)}{K_1(kR)}
  \frac{K_2(k'R)}{K_1(k'R)} \right|_{k' = p_1 - p_2 - k} .
\end{equation}
where $\sum p = p_1 - p_2 - \pout$
and $\dfour{\euclideank}$ is the four-dimensional Euclidean volume measure.

This diagram is difficult to evaluate for finite momenta $p_1$ and $p_2$,
so we shall work in an approximation where all external three-momenta
vanish: that is, $\vect{p_1} = \vect{p_2} = \vect{p}_{\stext{out}} = 0$.
This approximation matches Ref. \cite{doran-jackel}.
In this case $k^{\prime 2} = k^2$,
and the loop integral becomes somewhat more
tractable.  In addition, in our conventions,
the product of spinor coefficient functions $\bar{u}(p') u(p)$ evaluates to 2
when summed over spins at zero momentum.
Temporarily ignoring the various numerical pre-factors,
one has to evaluate the integral
\begin{equation}
  \label{eq:loop-integral}
  \int_{\mu}^{\Lambda} k \, \measure{k} \; \left( \frac{K_2(kR)}{K_1(kR)}
  \right)^2 .
\end{equation}
We have explicitly written in an upper cut-off at Euclidean momenta
$k \sim \Lambda$ and a lower cut-off at $k \sim \mu$.
This extremely simple regularization has the advantage that it is
easy to apply in the present context.

The Macdonald functions $K_\nu$ have asymptotics governed by
\begin{equation}
  K_\nu(z) \rightarrow
  \frac{\Gamma(\nu)}{2} \left( \frac{z}{2} \right)^{-\nu}
  \quad\ntext{as $z \rightarrow 0$}; \quad
  K_\nu(z) \rightarrow
  \sqrt{\frac{\pi}{2 z}} \e{-z} \quad \ntext{as $z \rightarrow +\infty$}.
\end{equation}
In the infra-red, aside from numerical factors, the ratio $K_2(kR) / K_1(kR)$,
behaves as a function of $k$ like $k^{-1}$.  Combining this behaviour with the
factor of $k^{-1}$
already present in the propagator
\eqref{eq:on-brane-prop}, one can see that any infra-red
divergence ought to be the same as in four dimensions.  However,
the large-$z$ asymptotics of $K_\nu(z)$ changes the
divergent behaviour in the ultra-violet.  To make an
estimate of \eqref{eq:loop-integral}, we write
$\int_\mu^\Lambda = \int_\mu^{1/R} + \int_{1/R}^\Lambda$ and approximate
the integrand using its asymptotic form in both regions (after
changing variable to $z=kR$),
\begin{equation}
  \label{eq:loop-divergences}
  \int_\mu^\Lambda k \, \measure{k} \; \left( \frac{K_2(kR)}{K_1(kR)}
  \right)^2 \approx \frac{4}{R^2} \int_{\mu R}^1 \frac{\measure{z}}{z}
  + \frac{1}{R^2} \int_1^{\Lambda R} z \; \measure{z} \sim
  - \frac{4}{R^2} \ln \mu R + \frac{1}{2} \Lambda^2 .
\end{equation}
We have discarded a term of order $\order{1/R^2}$, which should be a good
approximation provided $\Lambda^2 \gg 1/R^2$.  For an extra dimension
of order 1 mm, $R^{-1} \sim 1.97 \times 10^{-4} \; \ntext{eV}$, so
this is abundantly satisfied.
\eqref{eq:loop-divergences} lets us pick out the leading order divergence
in the ultra-violet and infra-red by making a small- or large-$k$
approximation in the integrand, as appropriate.

One sees that the ultra-violet divergence, which is logarithmic in
the four-dimensional case \cite{doran-jackel}, is modified to a
considerably worse quadratic divergence.  It is
natural to interpret this modification as due to interactions with
the Kaluza--Klein tower.
Despite this, the induced coupling remains proportional to
the classical quintessence potential $V(Q)$, so this correction
term does not destroy properties of the classical dynamics.
This is entirely analogous to the situation in four dimensions.

\subsection{Triangle diagram}
The amplitude for the second (`triangle') diagram of
Fig. \ref{fig:grav-couple} is
\begin{equation}
  \fl
  \atri = - \frac{m^2 \kappa^4}{\scale^2 R^2}
  \frac{V(\qclass)}{(2\pi)^{9/2}} \delta(\sum p) \int
  \dfour{\euclideank}
  \frac{\bar{u}(p_2)[ -i (\slashed{p}_1 - \slashed{k}) + m ] u(p_1)}
       {(p_1-k)^2 + m^2}
  \frac{1}{k} \frac{K_2(kR)}{K_1(kR)}
  \frac{1}{k'} \frac{K_2(k'R)}{K_1(k'R)} .
\end{equation}
which is to be evaluated at $k' = p_1 - p_2 - k$.
Using the Dirac equation, which says $-i \slashed{p}_1 u(p_1) = m u(p_1)$, the
numerator can be rewritten $\bar{u}(p_2)[i \slashed{k} + 2m] u(p_1)$.
At zero external momentum, $p_1 - p_2 \approx 0$ and
when summed over spins $\bar{u}(0) u(0) = 2$.  One can also approximate
$i k_a \bar{u}(0) \gamma^a u(0) = i k_0 \bar{u}(0) \gamma^0 u(0) = k_0$,
with a further factor of two arising from a sum over spins.  Therefore,
the numerator is just $2k_0 + 4m$, before Euclidean continuation.
At this point, one would typically complete the square in the
denominator and drop terms which are odd in $k_a$, because the
integral ought to be rotationally invariant.  However this procedure
is inconvenient in the present case, since one wishes to keep the
argument of the Macdonald functions simple.  Keeping track of factors
of $\imag$ gives
\begin{equation}
  \fl
  \atri = -16\pi\imag \frac{m^3 \kappa^4}{\scale^2 R^2}
  \frac{V(\qclass)}{(2\pi)^{9/2}} \delta(\sum p)
  \int_{\mu R}^{\Lambda R}\measure{z}\; z \left( \frac{K_2(z)}{K_1(z)} \right)^2
  \int_0^\pi \measure{\theta} \;
  \frac{\sin^2 \theta (1-\cos\theta)}{z^2+4m^2 R^2 \cos^2 \theta} ,
\end{equation}
where we have changed variable to $z = kR$.
This can be approximated using the same technique applied above for
the loop diagram.  We find,
\begin{eqnarray}
  \nonumber
  \fl \int_{\mu R}^{\Lambda R} \measure{z} \;
  z \left( \frac{K_2(z)}{K_1(z)} \right)^2 \int_0^\pi
  \measure{\theta} \; \frac{\sin^2 \theta (1-\cos\theta)}
  {z^2+4m^2 R^2 \cos^2 \theta} \sim &&
  \frac{\pi}{2} \ln \frac{\Lambda + \sqrt{\Lambda^2 + 4m^2}}
  {\overline{\Lambda}} \\ && + \frac{\pi}{m^2R^2} \ln
  \frac{\mu\left(1 + \sqrt{1+4m^2}\right)}{\mu + \sqrt{\mu^2 + 4m^2}} ,
\end{eqnarray}
where $\overline{\Lambda}$ is a reference energy scale of order
$m$. This exhibits a logarithmic divergence in both the infra-red and
the ultra-violet, but remains proportional to $V(\qclass)$
and so can be absorbed into a redefinition of the potential.
Just like the loop-diagram calculated above, it does not destroy
the classical potential.

This result is rather general.  Since these amplitudes couple only
to the quintessence potential through a vertex factor, which does not
change when moving four dimensional cosmology to the brane world,
the result is the same, even though the character of the divergences
has changed.
\subsection{Gravitational contribution to quintessence mass}
\begin{figure}
\begin{center}
\includegraphics[clip=true]{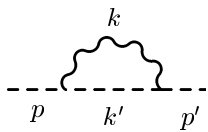}
\end{center}
\caption{Gravitational contribution to the quintessence self-energy.
The incoming and outgoing particles depicted by dashes are quintessence
particles.  The wavy line circulating in the loop is the graviton.\label{fig:grav-mass}}
\end{figure}

As a final application of the propagator \eqref{eq:on-brane-prop},
we suppose that the quintessence particle has some bare mass $m_Q$
and calculate the shift produced by the
contribution of the graviton loop in Fig.
\ref{fig:grav-mass}.
This vertices in the diagram take the form $(-1)m_Q^2 \delta_{ij}/4$
and come from the mass term $m_Q^2 Q^2/2$ in the potential and the
single-graviton coupling $\tr e/2$ to the quintessence field.
This diagram is a contribution to the self-energy
$\imag (2\pi)^4 \Pi^\ast(p)$ of the quintessence, at momentum $p$.  That gives,
with propagators for the external quintessence particles stripped off,
\begin{equation}
  \imag (2\pi)^4 \Pi^\ast(p) = - \frac{4\pi \imag m_Q^4 \kappa^2}{\scale R^2}
  \int_{\mu R}^{\Lambda R} \measure{z} \; \frac{K_2(z)}{K_1(z)}
  \int_0^\pi \measure{\theta} \; \frac{z^2 \sin^2\theta}{z^2+4m_Q^2 R^2 \cos^2 \theta}
\end{equation}
where we have aligned the polar axis with $p$, $m_Q$ is the
quintessence mass, we are assuming that
the particle is on shell, and we have made the standard change of variable $z = kR$.
With our conventions, the renormalized propagator becomes $\Delta'(p)
\propto (p^2 + m^2 - \Pi^\ast(p))^{-1}$, so that a negative
contribution to $\Pi^\ast$ gives a positive $\delta m^2$.
This gives the estimate
\begin{equation}
  \delta m_5^2
  \sim
  \frac{m_Q^4 \kappa^2}{4\pi^2} \frac{1}{\scale R^2} \left(
  - \frac{2}{m_Q R} \ln \frac{\mu}{2m_Q + \sqrt{\mu^2+4m_Q^2}} + \frac{R}{2} \Lambda
  \right) .
\end{equation}
That is, the divergence is linear in the ultra-violet and logarithmic
at infra-red.  Setting $m_Q$ to be currently of order the
Hubble rate, or $m_Q \sim 2.1 \times 10^{-33}h$ eV, and the infra-red
cutoff $\mu$ to the same,
we find
\begin{equation}
  \label{eq:mass-shift-estimate}
  \delta m_5^2 \sim 2.1 \times 10^{-141} h^3 \;
  \ntext{eV}^2 \quad \ntext{(5 dimensions)} .
\end{equation}
This is very small, and implies that interactions with new
gravitational physics associated with the brane world, such as the
Kaluza--Klein hierarchy and graviton transmission through the bulk,
do not seriously affect quintessence: its major problems remain
its couplings to and interactions with normal matter.  On the other hand,
the estimate \eqref{eq:mass-shift-estimate} is several orders of
magnitude smaller than a
comparable estimate for a four-dimensional cosmology with Planck scale
$\planckmass = 10^{19}$ GeV:
\begin{equation}
  \label{eq:fourd-mass-shift-estimate}
  \delta m_4^2 = \frac{1}{8\pi^2} \kappa_4^2
  m_Q^3 \Lambda \sim
  3.0 \times 10^{-127} h^3 \; \ntext{eV}^2 .
\end{equation}
In this case, there is no infra-red divergence so the magnitude of the
effect is controlled by the ultra-violet region.
The balance between \eqref{eq:mass-shift-estimate}
and \eqref{eq:fourd-mass-shift-estimate} is controlled by the ratio
\begin{equation}
  \frac{\delta m_5^2}{\delta m_4^2} = 4\alpha \frac{M_4^2}{M_5^3}
  \frac{\scale^2}{\Lambda} ,
\end{equation}
where $M_4$, $M_5$ are the four- and five-dimensional Planck scales,
respectively; $\Lambda$ is the four-dimensional ultra-violet cutoff;
$\scale$ is the five-dimensional AdS scale; and $\alpha = \ln (1+\surd 5)$
is a constant coming from the five-dimensional infra-red cutoff.
In the brane world, the interpretation of this difference involves
interactions with the Kaluza--Klein tower, whereas in the five-dimensional
picture one interprets the change as a result of the modified Planck
scale.
\section{Conclusions}
\label{sec:conclusions}
We have studied the constraints that arise on TeV-scale quintessence models 
from a variety of sources. Non-renormalisable operators in four dimensions are typically 
important implying that the quintessence potential needs to be computed 
from a higher-dimensional framework. This follows from the fundamental mismatch 
between the scale $M \sim \ntext{TeV}$
which determines the scale at which non-renormalisable operators
become important and the vacuum expectation value, $Q$, of the quintessence field 
which is typically of order the 4d Planck mass in tracker models. Perturbation theory
in $Q/M$ fails spectacularly. 

In contrast, the gravitational coupling of 
quintessence to fermionic matter in cosmologies of the Randall--Sundrum brane
world type does not yield significant constraints. This is easy to understand since
one expects the couplings of quintessence to ordinary
matter to be severely constrained and sensitive to the value of the
effective ultra-violet cut-off $\Lambda_{\ntext{uv}}$.  The brane world
significantly reduces the value of this cut-off, and so one would expect quite
radically different constraints on quintessence.

We find that one-loop effects introduce quantum corrections in the
effective potential just proportional to the classical potential $V$
and therefore can just be absorbed into a renormalization of $V$.
This is exactly the same as in the four-dimensional world and occurs
for the same reason: the vertices in the diagram generate factors of $V$,
not the propagator, and since this is the only quantity which changes
when one moves to the brane world the type and character of the
divergences one encounters changes, but the couplings remain the same.

We have also computed the lowest-order contribution from
graviton loops to the vacuum polarization of quintessence.  In this
case one must make a numerical estimate, and we find that the
brane universe typically induces a mass shift $\delta m^2$ very much
smaller than in four dimensions.  This shift is cut-off dependent,
and scales with the ratio of the four- and five-dimensional Planck scales,
the AdS curvature scale, and the inverse of the ultra-violet cutoff.
From the point of view of an observer on the brane,
we interpret this as the result of
interactions with the Kaluza--Klein hierarchy and with bulk gravitons.
The magnitude of this effect would render it undetectable
and in practice the dominant contributions to $\delta m^2$ would come
from matter fields on the brane.
\ack
DS is supported by a PPARC Studentship. BB is supported by a Royal 
Society/JSPS fellowship.
%
%
\appendix
\section{The gravitational propagator in the brane world}
\label{sec:grav-prop}
In this appendix we very briefly derive the graviton propagator in
the Randall--Sundrum scenario.  Since this calculation has already
appeared elsewhere \cite{giddings-katz} we omit details where they
coincide.  The authors
of Ref. \cite{giddings-katz} deduced the propagator by solving for the
gravitational Green's function whereas we employ the Fadeev--Popov
procedure, but naturally the final answers shall agree.
The principal result of this appendix is \eqref{eq:on-brane-prop}
for the on-brane propagator, which was used in the main text
for the diagram calculations in Section~\ref{sec:grav-couple}.

We adopt the conventional line element \cite{binetruy-deffayet-A,binetruy-deffayet-B},
for a given maximally symmetric three-metric $\gamma_{ij}$,
\begin{equation}
  \label{eq:metric}
  \measure{s}^2 = - n^2(t,y) \, \measure{t}^2 + a^2(t,y) \gamma_{ij} \,
  \measure{x}^i \, \measure{x}^j + \measure{y}^2 ,
\end{equation}
where $y$ is a Gaussian normal coordinate transverse to the brane.
This metric is taken to be a solution of the five-dimensional Einstein equations
with cosmological constant $\Lambda$ but vanishing bulk energy--momentum tensor.
The brane is considered to be imbedded at $y=0$, and there is
a $\zed{2}$ symmetry which acts on the Gaussian normal coordinate
as $y \mapsto -y$.  There is typically a coordinate horizon where the
Gaussian normal coordinates used in \eqref{eq:metric} break
down, and we write the location of this horizon as $y = \yh$.

Gravitational disturbances take the form of small perturbations
$h_{ab}$ to the metric: $\measure{s}^2 = (g_{ab} + h_{ab}) \,
\measure{x}^a \, \measure{x}^b$.  In a general $D$-dimensional spacetime,
$h_{ab}$ will transform as a representation of the isometry group $SO(1,D-1)$.
This describes the full degrees of freedom of the graviton.  Alternatively,
one could decompose $h_{ab}$ into its representations under the
brane isometry group, which consists of a tensor (in the
$\measure{x}^i \, \measure{x}^j$ sector of the metric) and supplementary
vector and scalar pieces (respectively,
for vectors, in the $\measure{t} \, \measure{x}^i$,
$\measure{y} \, \measure{x}^i$ sectors
and for scalars in $\measure{t}\, \measure{y}$,
$\measure{t}^2$ and $\measure{y}^2$ sectors) which must
be added in to complete the full degrees of freedom of the graviton.
In this paper we will deal only with the case of a flat, Minkowski brane
which possesses a larger isometry group and allows us to
re-absorb the vector and scalar pieces into the tensor perturbation.
For this reason, we only calculate the tensor propagator in this
section.

This piece of the perturbation is written $e_{ij}$ and takes the
metric form
\begin{equation}
  \measure{x}^i = - n^2(t,y) \, \measure{t}^2 + a^2(t,y) (\delta_{ij}
  + e_{ij}) \, \measure{x}^i \, \measure{x}^j + \measure{y}^2 .
\end{equation}
After integrating by parts and discarding surface terms,
the action is
\begin{equation}
  S = - \frac{1}{2\kappa^2} \int \dfive{x} \; n a^3 \; e^{ij} \left[
  2\delta_{i[j} \delta_{r]s} (\Box - \laplacian) - \frac{4}{a^2}
  \delta_{i[r} \partial_{j]} \partial_s \right] e^{rs}
\end{equation}
We have
introduced an operator $\Box$ describing $t$ and $y$ derivatives,
\begin{equation}
  \label{eq:graviton-action}
  \Box = \frac{1}{n^2} \ptwo{}{t} - \ptwo{}{y} + \frac{1}{n^2}
  \left( 3 \frac{\dot{a}}{a} - \frac{\dot{n}}{n} \right) \pd{}{t} -
  \left( 3 \frac{a'}{a} + \frac{n'}{n} \right) \pd{}{y}
\end{equation}
and $\laplacian$ is the $\delta_{ij}$ Laplacian.  The brane isometry
group $\iso$ now appears as an invariance of this action:
\eqref{eq:graviton-action} is
invariant under $\iso$ transformations (spatial rotations and translations).

One passes to the quantum theory by defining correlation functions of
the field $e_{ij}$ using the functional integral prescription,
$\langle e_{ij}(x) \cdots
e_{mn}(y) \rangle = \int \fmeasure{e_{rs}} \, e_{ij}(x) \cdots
e_{mn}(y) \; \exp{\imag S[e_{rs}]}$.  To render this integral (formally) finite,
one divides out by the volume of the gauge group.  This is the
Fadeev--Popov procedure \cite{fadeev-popov}, amounting to the inclusion
of an extra gauge-fixing contribution in the action,
\begin{equation}
  S \mapsto S + \frac{1}{2\kappa^2} \int \dfive{x} \; na^3 \;
  \frac{1}{2\xi} ( \partial e_b - \alpha
  \partial_b e)^2
\end{equation}
where $\xi$ and $\alpha$ are arbitrary numbers.  We have suppressed
irrelevant indices by writing $\partial e_b = \partial^a e_{ab}$ and
$e = \tr e_{ab}$.  Notice that in these formulas we are considering
$e_{ab}$ to be a full five-dimensional tensor which is zero on
$t$ or $y$ indices.  Where derivatives
are contracted with $e_{ab}$ this makes no difference ($\partial^a e_{ab}
= \partial^i e_{ib}$), but where two derivatives become contracted with
themselves as in $\partial^a \partial_a$ one must include contributions
from the $t$ and $y$ sectors.

After integrating by parts one obtains
the gauge-fixed action,
\begin{eqnarray}
  \fl\nonumber
  S = - \frac{1}{2\kappa^2} \int \dfive{x} \; n a^3 \; e^{ij} \Bigg\{ & &
  \left[ \Box - \frac{\laplacian}{a^2} \right] \left[
  - \frac{1}{4} \delta_{ir} \delta_{js} +
  \delta_{ij} \delta_{rs} \left( \frac{1}{4} - \frac{\alpha^2}{2\xi} \right)
  \right] \\
  & & - \frac{1}{a^2} \delta_{ir} \partial_j \partial_s \left(
  \frac{1}{2} - \frac{1}{2\xi} \right) + \frac{1}{a^2}
  \delta_{ij} \partial_r \partial_s \left( \frac{1}{2} - \frac{\alpha}{\xi}
  \right) \Bigg\} e^{rs} .
\end{eqnarray}
In order to simplify this result, it is convenient to set $\xi = 1$,
$\alpha = 1/2$ in which case both terms involving $\alpha$ disappear
and one is left with the reduced action
\begin{equation}
  S = - \frac{1}{2} \int \dfive{x} \; na^3 \; e^{ij} D_{ijrs} e^{rs}
\end{equation}
where $D_{ijrs}$ is the operator
\begin{equation}
  \label{eq:dop-definition}
  D_{ijrs} = \frac{1}{4\kappa^2} \left( 2 \delta_{i(r}\delta_{s)j} -
  \frac{1}{2} \delta_{ij} \delta_{rs} \right) \left( \Box -
  \frac{\laplacian}{a^2} \right) .
\end{equation}
These choices for $\alpha$ and $\xi$ coincide with the four-dimensional
case.
The propagator $\Delta^{rsmn}(x,y)$ satisfies
\begin{equation}
  \label{eq:define-propagator}
  D_{ijrs} \Delta^{rsmn}(x,y) = -\imag \delta^5(x-y) \delta^m_{(i} \delta^n_{j)}
\end{equation}
where $\delta^5$ is the covariant $\delta$-function.

In a quite general homogeneous brane world, the metric functions $n$ and $a$ are
not equal and both depend on $t$ and $y$, so one has the
three Killing vectors $\partial / \partial x^i$ which generate translations along
the spacelike coordinate axes, but no other translational Killing vectors.
For this reason, it is sensible to try and
diagonalize $\Delta^{rsmn}$
as a Fourier transform in the $x^i$, but one will not be able
to deal with the $t$ and $y$ dependence in the same way.  Therefore,
\begin{equation}
  \label{eq:proposed-propagator}
  \Delta^{rsmn} = -\imag \rho^{rsmn} \int
  \frac{\dthree{p}}{(2\pi)^3} \; \e{-\imag \vect{p} \cdot (\vect{x} - \vect{y})}
  G(\vect{p}; x_0, y_0; x_5, y_5) ,
\end{equation}
where $\rho^{rsmn}$ is the combination
\begin{equation}
  \label{eq:index-inversion}
  \rho^{rsmn} = \delta^{r(m} \delta^{n)s} - \frac{1}{d-2} \delta^{rs}
  \delta^{mn} ,
\end{equation}
and $d = \delta^i_i$ is the trace of the $\iso$ Kronecker delta.
We are now adopting a convention of writing the
coordinates of any point $x$ in spacetime as $(x^0, \vect{x}, x^5)$.
Substituting \eqref{eq:proposed-propagator} into
\eqref{eq:define-propagator} shows that $G$ must obey the equation
\begin{equation}
  \left( \Box + \frac{\vect{p}^2}{a^2} \right) G =
  2 \kappa^2 \delta(x^0 - y^0) \delta(x^5 - y^5) .
\end{equation}

\subsection{The Randall--Sundrum propagator}
At this point, one can make no further progress without specifying some
explicit form for $a$ and $n$.  The simplest choice is to take the
brane to be empty of matter, except for some intrinsic tension
\cite{randall-sundrum-A} which is tuned to give a Minkowski brane.
The line element is
\begin{equation}
  \measure{s}^2 = \e{-2\scale |y|} ( - \measure{t}^2 +
  \delta_{ij} \, \measure{x}^i \, \measure{x}^j ) + \measure{y}^2 .
\end{equation}
In this special case, the functions $a$ and $n$ do turn out to be equal,
and all quantities are independent of the cosmic time $t$, so
one recovers $\partial / \partial t$ as a Killing symmetry.  This is
a great convenience, because one can now write $G$ as a Fourier integral
$G = \int \measure{\omega} \, (2\pi)^{-1} \e{\imag \omega (x^0 - y^0)}
\tilde{G}$
in $x^0 - y^0$, leaving only an ordinary differential equation
for the $x^5$, $y^5$ dependence of the Fourier transform
$\tilde{G}$.  Changing to a conformal bulk coordinate $z$ defined by
$\measure{x}^5 = a \, \measure{x}^z$, this ordinary differential
equation turns out to be just the Bessel equation,
\begin{equation}
  \label{eq:rs-propagator-eqn}
  \bop{} \tilde{G} =
  - 2 \kappa^2 \frac{y^z}{\scale (x^z)^4} \delta(x^z - y^z) ,
\end{equation}
where $\bop{}$ is the Bessel operator,
\begin{equation}
  \bop{} = \ptwo{}{(x^z)} + \frac{1}{x^z} \pd{}{x^z} + \left(
    \beta^2 - \frac{4}{(x^z)^2} \right) .
\end{equation}
In these coordinates, the location of the brane is $z = \scale^{-1}$,
but we will often take its location to be arbitrary and write
$z = R$ instead.  When making numerical estimates, we restore
$\scale R = 1$.
We have introduced a new quantity $\beta$ defined by $\beta^2
= \omega^2 - \vect{p}^2$ and define a four-vector $k=(\omega,\vect{p})$
with $k^2 = - \beta^2$ in our signature $\diag(-1,1,1,1)$.

From this point, out derivation coincides with the earlier derivation
of Giddings, Katz \& Randall \cite{giddings-katz}, so we omit further
details and merely state the result.  The general solution for
$\tilde{G}$ is a combination of Bessel functions, and by integrating
\eqref{eq:rs-propagator-eqn} over a small neighbourhood of
$x^z = y^z$ one obtains a continuity condition on $\tilde{G}$
and a step condition on $\partial\tilde{G}/\partial x^z$.
Demanding that the normal derivative of $G$ vanish at the brane,
and that positive frequency waves be purely ingoing in the far-field,
together with the junction conditions at the brane 
allows one to solve uniquely for $G$.  The result for the entire
propagator is
\begin{equation}
\label{eq:graviton-propagator}
  \Delta^{rsmn} = \rho^{rsmn} \int \frac{\dthree{p} \,
  \measure{\omega}}{(2\pi)^4} \e{-\imag \vect{p}\cdot(\vect{x} - \vect{y})
  + \imag \omega (x^0 - y^0)} W(x^z,y^z;R) ,
\end{equation}
where $W(x^z,y^z;R)$ satisfies
\begin{equation}
  \fl
  W( x^z,y^z;R) = -\imag \left( \frac{x^z}{y^z} \right)^2
  \frac{\kappa^2 \pi}{\scale} \frac{H_2^{(1)}(\beta z_>)}
  {H_1^{(1)}(\beta R)} \left( J_1(\beta R) H_2^{(1)}(\beta z_<) -
  H_1^{(1)}(\beta R) J_2(\beta z_>) \right)
\end{equation}
in which $z = R$ is the location of the brane
and $z_<$, $z_>$ are respectively $\min \{ x^z, y^z \}$, $\max
\{ x^z, y^z \}$.  $H_\nu^{(1)}$ is the Hankel function of order $\nu$ of the
first kind.  This propagator agrees with the expressions
(2.14)--(2.15) given in Ref. \cite{giddings-katz} once
notational differences have been taken into account.

There is an obvious simplification of \eqref{eq:graviton-propagator}
in the special case that both endpoints $x$, $y$ are taken on the
brane.  One finds,
\begin{equation}
  \label{eq:on-brane-prop}
  \left .\Delta^{rsmn}\right|_{\stext{on-brane}} =
  \frac{2 \kappa^2}{\scale R} \rho^{rsmn} \int
  \frac{\dfour{k}}{(2\pi)^4} \e{-\imag k \cdot (x-y)} \frac{1}{k}
  \frac{K_2(kR)}{K_1(kR)}
\end{equation}
where $K_\nu$ is the Macdonald or Basset function
\begin{equation}
  K_\nu(z) = \frac{1}{2} \pi \imag^{\nu + 1} H_\nu^{(1)}(\imag z)
\end{equation}
and is entirely real.

In the text, we make use of Wick rotation to Euclidean signature.
Since the graviton propagator has acquired a much more complicated
space-time dependence than the propagators of Minkowski space
fields, there may be more obstructions to this procedure than can be
dealt with by moving the $1/k$ poles of \eqref{eq:on-brane-prop}
off-axis.  In particular, if the Macdonald functions $K_\nu$ have
singularities anywhere on $\complex{}$, then a Wick rotation could not be justified.
To see that this is not so, it is convenient to make use of the
following integral representation of the Macdonald function
\begin{equation}
  K_{\nu}(z) = \sqrt{\frac{\pi}{2z}} \frac{\e{-z}}{\Gamma(\nu + 1/2)}
  \int_0^\infty \e{-t} t^{\nu - 1/2} \left(1 - \frac{t}{2z} \right)^{
  n-1/2} \; \measure{t}
\end{equation}
From this it is clear that $K_{\nu}$ has a singularity
at $z=0$ for any $\nu$ but is otherwise analytic everywhere.
Moreover it is not zero except at $z = \infty$.  There is an
essential singularity at $\infty$ stemming from the exponential,
but since the contour remains fixed at $z=0$ and $z=\infty$, this
does not interfere with the analytic continuation.
%
%
\providecommand{\bysame}{\leavevmode\hbox to3em{\hrulefill}\thinspace}
\providecommand{\MR}{\relax\ifhmode\unskip\space\fi MR }
\providecommand{\MRhref}[2]{%
  \href{http://www.ams.org/mathscinet-getitem?mr=#1}{#2}
}
\providecommand{\href}[2]{#2}
\hspace{1cm}
\begin{flushleft}\large\textbf{References}\end{flushleft}

\end{document}